\documentclass[hyper,12pt,letterpaper]{JHEP3} 

\usepackage{amsmath,amsfonts,amssymb}
\usepackage{graphicx}
\usepackage{young}

\numberwithin{equation}{section}

\def\cN{\mathcal{N}}
\def\cM{\mathcal{M}}
\def\cO{\mathcal{O}}

\def\cD{\mathcal{D}}
\def\cF{\mathcal{F}}

\def\cJ{\mathcal{J}}
\def\cK{\mathcal{K}}
\def\cP{\mathcal{P}}

\def\cH{\mathcal{H}}
\def\cW{\mathcal{W}}
\def\cV{\mathcal{V}}
\def\bV{\mathbb{V}}

\def\bR{\mathbb{R}}
\def\bC{\mathbb{C}}
\def\bk{\mathbf{k}}
\def\CP{\mathbb{CP}}
\def\SU{SU}
\def\SO{SO}
\def\SL{SL}

\def\U{U}
\def\Ext{\mathop{\mathrm{Ext}}\nolimits}
\def\tr{\mathop{\mathrm{tr}}\nolimits}
\def\diag{\mathop{\mathrm{diag}}}

\def\vev#1{\langle#1\rangle}
\def\vvevv#1{\langle\!\langle#1\rangle\!\rangle}
\def\kket#1{|#1\rangle\!\rangle}
\def\ket#1{|#1\rangle}
\def\bbra#1{\langle\!\langle#1|}

\let\tilde\widetilde
\let\eth\lambda

\title{Affine SL(2) conformal blocks \\
from 4d gauge theories}

\author{Luis F. Alday and Yuji Tachikawa\\

{\tt alday,yujitach@ias.edu}\\
School of Natural Sciences, Institute for Advanced Study, \\
Princeton, New Jersey 08540, USA
}

\abstract{
We study Nekrasov's instanton partition function of four-dimensional $\cN=2$ gauge theories in the presence of surface operators. This can be computed order by order in the instanton expansion by using results available in the mathematical literature.
Focusing in the case of $SU(2)$ quiver gauge theories, we find that the results agree with a modified version of the conformal blocks of affine $SL(2)$   algebra. 
These conformal blocks provide, in the critical limit, the eigenfunctions of the corresponding quantized Hitchin Hamiltonians.
}

\begin{document}

\tableofcontents

\section{Introduction}
Last year, a systematic method to construct four-dimensional $\cN=2$ theories by wrapping $N$ M5-branes on a Riemann surface with punctures was developed in \cite{Gaiotto:2009we}.
This construction suggested that physical quantities of the resulting four-dimensional theory should be closely related to those of the theory on the two-dimensional Riemann surface.
Indeed, it was found in \cite{Alday:2009aq} that the instanton partition function \cite{Nekrasov:2002qd} of the four-dimensional theory obtained from two M5-branes is equal to the conformal block of the Virasoro algebra associated to the Riemann surface.  The central charge is given by \begin{equation}
c=1+6\frac{(\epsilon_1+\epsilon_2)^2}{\epsilon_1\epsilon_2}
\end{equation} where $\epsilon_1$  and $\epsilon_2$ are the two rotation angles which enter the definition of the instanton partition function.
This observation was soon extended to the case with more than two M5-branes in \cite{Wyllard:2009hg,Mironov:2009by}.

The six-dimensional theory on the M5-branes has two types of defects: those with two-dimensional worldvolume and those with four-dimensional worldvolume. The former are ends of the M2-branes which have boundaries on the M5-branes, and the latter are the intersections of the original stack of M5-branes with another set of M5-branes.
When M5-branes are put on four-dimensional Minkowski space $X_4$ times a Riemann surface $C_2$, we have different choices for distributing the worldvolume dimensionality of a given defect between $X_4$ and $C_2$, see Table~\ref{table}.

\begin{table}[h]
\begin{center}
\begin{tabular}{r|c|cc|l}
&6d & $X_4$ & $C_2$ & comments \\
\hline
i)&2d & 2 & 0  & Surfaces on $X_4$\\
ii)&2d & 1 & 1  & Loops on $X_4$ \\
\hline
iii)&4d & 4 & 0  & Changes theory on $X_4$ \\
iv)&4d & 3 & 1  & Domain wall on $X_4$ \\
v)&4d & 2 & 2  & Surfaces on $X_4$. Changes theory on $C_2$
\end{tabular}
\end{center}
\caption{Various type of defects of six-dimensional theory.\label{table}}
\end{table}

Case i) corresponds to a two-dimensional defect on a product of a surface in $X_4$ and a point in $C_2$. One finds that the degenerate fields of the two-dimensional theory give a surface operator in $X_4$ \cite{Alday:2009fs,Kozcaz:2010af}.
Case ii) corresponds to wrapping a two-dimensional defect on a product of a loop in $X_4$ and a loop in $C_2$.
This construction led to the classification of the loop operators in four dimensions in terms of the loops on $C_2$ \cite{Drukker:2009tz} and the evaluation of their expectation values \cite{Alday:2009fs,Drukker:2009id}.
Case iii) corresponds to a four-dimensional defect  filling $X_4$ and placed  at a point on $C_2$. This changes the four-dimensional theory on $X_4$ \cite{Gaiotto:2009we}.
Case iv) corresponds to wrapping a four-dimensional defect on a loop in $C_2$ to construct a domain wall in $X_4$. This was studied in \cite{Drukker:2010jp}.

The subject of this paper is the case v) where we wrap the four-dimensional defect on a product of a surface in $X_4$ times the whole of $C_2$.
This gives a surface operator on $X_4$, and changes the theory living on $C_2$.

A basic type of four-dimensional defects of the theory on $N$ M5-branes is labeled by a Young diagram with $N$ boxes \cite{Gaiotto:2009we}.
Therefore the surface operator coming from this construction should also be labeled in the same way.
For $\cN=4$ $\SU(N)$ theory, i.e.~when the Riemann surface $C_2$ is a torus without any puncture,  surface operators with exactly this property were studied in detail by \cite{Gukov:2006jk}.
There, the Young diagram dictates how the gauge field diverges close to the surface operator: \begin{equation}
A_\mu dx^\mu \sim \diag(\alpha_1,\alpha_2,\ldots,\alpha_N) i d\theta
\end{equation} where $\theta$ is the angular coordinate of the transverse plane to the surface operator.
Then the Young diagram $N=N_1+N_2+\cdots+N_s$ corresponds to the choice of $\alpha_i$ where $\diag(\alpha_1,\ldots,\alpha_N)$ commutes with $S[\U(N_1)\times  \cdots \times \U(N_s)]$.
In this paper we will only consider the surface operator corresponding to the most generic choice of $\alpha_i$, where the commutant is $\U(1)^{N-1}$.  We call this a \emph{full} surface operator.
Note that when there are multiple gauge groups in the four-dimensional gauge theory, this surface operator creates singularities in all the corresponding gauge fields.

Mathematicians have developed a method to obtain the instanton partition function in the presence of the full surface operator \cite{Braverman:2004vv,Braverman:2004cr,Negut,FFNR}.
In fact it was already proved in 2004 that the instanton partition function in the presence of the full surface operator has in it the structure of affine $\SL(N)$ algebra whose level is given by \begin{equation}
k=-N-\frac{\epsilon_2}{\epsilon_1}.
\end{equation}
Our first objective in this paper is to explain, in terms palpable to string theorists,  what mathematicians already know concerning this setup. We will then generalize it slightly.
Our concrete calculation focuses on the simplest case $N=2$.
We explicitly demonstrate that the instanton partition functions in the presence of the full surface operator is equal to the conformal blocks of $\SL(2)$ affine algebra with a certain small modification.

Our second objective is to study the limit $\epsilon_2\to 0$, which was also the limit studied by Nekrasov and Shatashvili \cite{Nekrasov:2009rc}.
There, it was found that, in this limit, the instanton partition function \emph{without} the surface operator determines the eigen\emph{values} of the quantized Hamiltonians of the integrable systems associated to the four-dimensional theory.
Here, we will see that the partition function \emph{with} the surface operator, which is basically the conformal block of the affine Lie algebra, in this limit provides the simultaneous eigen\emph{function} of the quantized Hamiltonians.
Therefore, the full surface operator provides a way to bridge the results of \cite{Alday:2009aq} and the results of \cite{Nekrasov:2009rc}.

The rest of the paper is organized as follows. We start in Sec.~\ref{without} by recalling how  the instanton partition function without the surface operator is calculated. We recast the partition function in a form that suggests the equivalence to a conformal block.
In Sec.~\ref{with}, we introduce  the surface operator. We write down the instanton partition function in this setup, using results available in the mathematical literature.
We compute the partition function explicitly for $\SU(2)$ quiver theories and find that it coincides with a modified version of the conformal blocks of affine $\SL(2)$ algebra.
In Sec.~\ref{eigen}, we consider the critical limit, i.e. the limit $\epsilon_2/\epsilon_1\to 0$,
of such conformal blocks. We argue that we can recover from them the eigenfunctions of the corresponding Hitchin Hamiltonians. We explicitly consider the case of the one-point function on the torus, which corresponds to the elliptic Calogero-Moser Hamiltonian.
In Sec.~\ref{conclusions}, we conclude and defer several technical details to the appendices.

While preparing this manuscript, \cite{Teschner:2010je} appeared, which overlaps with Sec.~\ref{eigen} of this paper.

\section{Instanton partition function without surface operator}\label{without}
We start  our discussion by recalling the definition of Nekrasov's partition function of an $\cN=2$ gauge theory and the way to calculate it.
We then show how to reformulate the partition function so that its similarity to the conformal blocks is manifest.
The system we consider in this section does not have surface operators in it. However, the  reformulation of the partition function will be crucial in the next section, where we make the connection with the computation in presence of surface operators done by mathematicians.

\subsection{Definition}
Nekrasov's partition function of an $\cN=2$ theory on $\bR^4$  is defined by the following procedure:
one introduces two deformation parameters $\epsilon_{1,2}$ which are associated to  rotations on  the $x^1$--$x^2$ plane and the $x^3$--$x^4$ plane, respectively. In particular $\epsilon_1$ breaks the translational invariance in the directions $x^1$ and $x^2$,
and likewise $\epsilon_2$ does in the directions $x^3$ and $x^4$.
After introducing such deformations it makes sense to talk about the supersymmetric partition function of the system, because the system is effectively compactified to the region close to the origin of $\bR^4$.
Let us denote the partition function by $Z_\text{full}(\epsilon_1,\epsilon_2; a; m)$, where $a$ stands for  the vacuum expectation values of the vector multiplets and $m$ stands for the hypermultiplet masses.
The partition function is known to behave,  in the limit $\epsilon_{1,2}\ll a,m$, as
\begin{equation}
Z_\text{full}(\epsilon_1,\epsilon_2;a;m) \sim \exp(-\frac{F(a;m)}{\epsilon_1\epsilon_2}+ \cdots )\label{limit-to-prepotential}
\end{equation} where $F(a,m)$ is the prepotential of the system.

When $Z$ is calculated in the weak-coupling regime,
it receives three distinct contributions, namely,  classical, one-loop and instanton contributions: \begin{equation}
Z_\text{full}(\epsilon_1,\epsilon_2; a ; m )=Z_\text{classical}
Z_\text{1-loop} Z_\text{inst}.
\end{equation} In the following we mainly talk about the instanton contribution;
$Z$ will stand for $Z_\text{inst}$ unless otherwise mentioned.

Let us consider $\cN=2$ $\SU(N)$ theory with a massive adjoint hypermultiplet as an example to recall how the calculation is performed.
The instanton partition function receives contributions only from half-BPS configurations. The gauge field strength then needs to be anti-self-dual. Let us denote by $\cM_{N,k}$ the moduli space of anti-self-dual instanton configurations with instanton number $k$. This is a hyperk\"ahler manifold of real dimension $4Nk$.
Then \begin{equation}
Z_\text{inst}(\epsilon_1,\epsilon_2; a; m)
= \sum_k q^k \int_{\cM_{N,k}}
\omega(\epsilon_i,a,m) \label{integral}
\end{equation} where $\omega(\epsilon,a,m)$ is a certain differential form on the instanton moduli which is determined by the matter content of the theory.

$\cM_{N,k}$ is a highly singular space because of point-like instantons.
These singularities make the evaluation of \eqref{integral} not straightforward.
To ameliorate the situation, we replace $\cM_{N,k}$ in the integral \eqref{integral} by a different space  $\tilde \cM_{N,k}$ which is a blowup of $\cM_{N,k}$.
$\tilde \cM_{N,k}$ is simultaneously the moduli space of  $\U(N)$ $k$-instanton configurations on non-commutative  $\bR^4$ and of rank-$N$ torsion-free sheaves on $\CP^2$ framed at infinity.

The integrand \eqref{integral} is such that one can use the fixed point theorem\begin{equation}
Z_\text{inst}(\epsilon_1,\epsilon_2;a;m)
=\sum_k q^k \sum_{p: \ \text{fixed points on $\tilde\cM_{N,k}$}}
 \frac{z_\text{hyper}[p](\epsilon_i;a;m)}{z_\text{vector}[p](\epsilon_i;a)}.\label{sum}
\end{equation}
Here $p$ runs over the fixed points of the action
$\U(1)^2\times \U(1)^N\subset \SO(4)\times \U(N)$
where $\SO(4)$ acts on $\tilde\cM_{N,k}$ by rotating the spacetime\footnote{Strictly speaking, the regularization by the noncommutativity parameter or by translating to the sheaves on $\CP^2$ breaks the spacetime symmetry down to $\U(2)$.}
$\bR^4$  and $\U(N)$ by a global gauge rotation.
The numerator and the denominator in the summand in \eqref{sum} come from the one-loop fluctuation around $p$, the fixed point configuration of the instantons. As usual, the vector multiplet behaves like a boson, thus contributing to the denominator,
while the hypermultiplet behaves like a fermion, hence contributing to the numerator.

A fixed point $p\in \tilde \cM_{N,k}$ is labeled by an $N$-tuple of Young diagrams $\vec Y = (Y_1,\ldots,Y_N)$ such that the total number of boxes $|\vec Y|=\sum |Y_i|$ equals $k$. Each of $Y_i$ specifies a $\U(1)$ non-commutative instanton of instanton number $|Y_i|$, or equivalently a rank-$1$ torsion-free sheaf.
The configuration $\vec Y$ is then given by a direct sum of these $\U(1)$ configurations.
Therefore in the following we use the symbol $\vec Y$ instead of $p$ to label the fixed points. The explicit form of $z_\text{vector,hyper}[\vec Y]$ was  determined in \cite{Nekrasov:2002qd,Flume:2002az,Nakajima:2003pg,Fucito:2004gi} and given, for instance, in appendix B to \cite{Alday:2009aq}.

\subsection{Reformulation}\label{reformulation}
There is a nice way to reformulate the instanton partition function \eqref{sum} which makes manifest its similarity to the conformal blocks.
For illustration, it is better to consider a slightly more complicated gauge theory.
So let us consider a theory with gauge group  $\SU(N)_1\times \SU(N)_2\times \SU(N)_3$ and bifundamental hypermultiplets charged under $\SU(N)_{i}\times \SU(N)_{j}$ for $(i,j)=(1,2)$, $(2,3)$ and $(3,1)$.
Then the instanton partition function is given by \begin{multline}
Z(\vec a_1,\vec a_2,\vec a_3;m_1,m_2,m_3)=
\sum_{k_1,k_2,k_3} q_1^{k_1}q_2^{k_2}q_3^{k_3}
\sum_{\vec Y_1,\vec Y_2,\vec Y_3}\\
\frac{z_\text{bif}[\vec Y_1,\vec Y_2](\vec a_1,\vec a_2;m_1)
z_\text{bif}[\vec Y_2,\vec Y_3](\vec a_2,\vec a_3;m_2)
z_\text{bif}[\vec Y_3,\vec Y_1](\vec a_3,\vec a_1;m_3)}
{z_\text{vec}[\vec Y_1](\vec a_1)z_\text{vec}[\vec Y_2](\vec a_2)z_\text{vec}[\vec Y_3](\vec a_3)}.\label{complicated}
\end{multline}
Here, $\vec a_i$ and $\vec Y_i$ specify
 the vacuum expectation value and the fixed configuration of $\SU(N)_i$, respectively.
 $z_\text{bif}[\vec Y_1,\vec Y_2](\vec a_1,\vec a_2;m_1)$ is the contribution of a bifudamental hypermultiplet charged under $\SU(N)_1\times \SU(N)_2$.

Introduce a vector space $\cH_{\vec a}$ whose basis is $\kket{\vec Y}_{\vec a}$ for all $N$-tuple  $\vec Y$ of Young diagrams.
We introduce an inner product on $\cH_{\vec a}$,\begin{equation}
_{\vec a}\vvevv{\vec Y|\vec Y'}_{\vec a}= \delta_{\vec Y,\vec Y'} /{z_\text{vector}[\vec Y](\vec a)}.\label{innerproduct}
\end{equation} Note that $\cH_{\vec a}$ and $\cH_{\vec a'}$ might look similar but have a different inner product.\footnote{
This space $\cH_{\vec a}$ is a natural mathematical object: \[
\cH_{\vec a}=\bigoplus_k
H_{\U(1)^2\times \U(1)^N}(\tilde\cM_{N,k})
\otimes \bC(\epsilon_{1},\epsilon_2,a_1,\ldots,a_N),
\] where $H_{G}(X)$ is the equivariant cohomology of $X$ with respect to the action of the group $G$.
Here we identified $H_{\U(1)^2\times \U(1)^N}(\textrm{pt})=\bC[\epsilon_1,\epsilon_2,a_1,\ldots,a_N]$.
Then the inner product \eqref{innerproduct} is exactly the equivariant integral over $\tilde\cM_{N,k}$ of two elements in the equivariant cohomology. For more on this point, see e.g.~\cite{Negut,Nakajima:2003uh}. } A bifundamental hypermultiplet defines a map \begin{equation}
\Phi_{\vec a\, m\, \vec b} : \cH_{\vec b} \to \cH_{\vec a}
\end{equation}  whose matrix elements are given by \begin{equation}
\Phi_{\vec a\, m\, \vec b}\kket{\vec Y}_{\vec b}=\sum_{\vec Y'}\frac{z_\text{bif}[\vec Y',\vec Y](\vec a,\vec b,m)}
{z_\text{vector}[\vec Y](\vec b)}\kket{\vec Y'}_{\vec a}.
\end{equation}
Let us also define an operator ${\bf N}$ on $\cH_{\vec a}$ which counts the number of boxes: \begin{equation}
{\bf N}\kket{\vec Y}_{\vec a}=|\vec Y|\, \kket{\vec Y}_{\vec a}.
\end{equation}

We further define $\Phi_m(z)$ by \begin{equation}
\Phi_{\vec a,m,\vec b}(z)= z^{\bf N} \Phi_{\vec a,m,\vec b} z^{-\bf N}.
\end{equation}
Then the instanton partition function \eqref{complicated}
can be recast into a trace of operators \begin{equation}
Z=\tr_{\cH_{\vec a_1}}
(q_1q_2q_3)^{\bf N}
\Phi_{\vec a_1\, m_1\, \vec a_2}(z_1)\,
\Phi_{\vec a_2\, m_2\, \vec a_3}(z_2)\,
\Phi_{\vec a_3\, m_3\, \vec a_1}(z_3)\,
\end{equation} where \begin{equation}
z_1=q_1,\quad z_2=q_1q_2,\quad z_3=q_1q_2q_3.
\end{equation}

This interpretation was put forward for the case of $\U(1)$ quivers
by Carlsson and Okounkov \cite{CarlssonOkounkov}.
In that case the dependence on the parameter $\vec a$ drops out.
The space $\cH$ is identified with a free-boson Fock space $V_\text{Fock}$ and the operator $\Phi_m$ can be shown to be equal to a certain vertex operator: \begin{equation}
\Phi_m(z)= V_{\epsilon_1+\epsilon_2-m,m}(z)
\end{equation}
where
\begin{equation}
V_{\alpha,\beta} (z)=  \exp\left[\frac{\alpha}{\sqrt{\epsilon_1\epsilon_2}}\varphi_-(z)\right]
\exp\left[\frac{\beta}{\sqrt{\epsilon_1\epsilon_2}} \varphi_+(z)\right].
\end{equation}
Here $\varphi(z)$ is a free-boson field with the OPE $\varphi(z)\varphi(0)\sim -\log z$, and $\varphi(z)=\varphi_+(z)+\varphi_-(z)$ is its decomposition into positive and negative modes.
Note that $V_{\alpha,\beta}$ is a standard normal-ordered exponential when $\alpha=\beta$.

In \cite{Alday:2009aq,Wyllard:2009hg,Mironov:2009by} the situation with $N=2,3$ was studied. There, $Z$ was found to be equal to a certain ``$\U(1)$ factor'' times the conformal block of $W_N$ algebra.
This $\U(1)$ factor is almost the same as the partition function of $\U(1)$ quivers, as noticed by \cite{Wyllard:2009hg}, and can be expressed in terms of the vertex operator $V_{\alpha,\beta}$ defined above.

Combining this set of information, one can reformulate the relation between instanton partition functions and conformal blocks succinctly as follows:
\begin{enumerate}
\item  The space $\cH_{\vec a}$ introduced above can be decomposed as \begin{equation}
\cH_{\vec a} = V_\text{Fock} \otimes \cV_{\vec\alpha(\vec a)}
\qquad\text{where}\quad \vec\alpha(\vec a)=\frac{\vec a}{\sqrt{\epsilon_1\epsilon_2}} +Q\vec\rho.
\label{decomposition}
\end{equation}
$\cV_{\vec\alpha}$ is the Verma module of $W_N$ algebra
whose central charge is \begin{equation}
c=(N-1)+(N-1)N(N+1)Q^2 \qquad
\text{with}\quad Q^2=\frac{(\epsilon_1+\epsilon_2)^2}{\epsilon_1\epsilon_2}
\end{equation}  labeled by the momenta $\vec\alpha$.\footnote{Here, the momentum refers to  that of the free-field representation of $W_N$ algebra. See Appendix~\ref{free}.}
In particular, when $N=2$ the W-algebra is the Virasoro algebra, and $V_{\vec\alpha}\equiv \cV_{(\alpha,-\alpha)}$ is the Verma module generated by  the primary state with dimension  $\alpha(Q-\alpha)$.
\item The map $\Phi_{\vec a,m,\vec b}:\cH_{\vec b}\to \cH_{\vec a}$ determined by the bifundamental hypermultiplet decomposes under \eqref{decomposition} as the tensor product of the vertex operator of Carlsson and Okounkov and the insertion of a primary field $\cV_{N\mu\vec \chi}$ of $W_N$ algebra:
\begin{equation}
\Phi_{\vec a,m,\vec b}=V_{\epsilon_1+\epsilon_2-m,Nm} \otimes \cV_{N\mu\vec \chi}:
V_\text{Fock}\otimes \cV_{\vec\alpha(\vec b)}
\to
V_\text{Fock}\otimes \cV_{\vec\alpha(\vec a)}
\end{equation}
where $\mu=m/\sqrt{\epsilon_1\epsilon_2}$ and
$\vec\chi$ is  the first weight vector of $\SU(N)$ as determined in \cite{Wyllard:2009hg}.
For $N=2$, $\cV_{N\mu\vec\chi}$ is the insertion of the primary state
with dimension $\mu(Q-\mu)$.
\end{enumerate}

There are already many pieces of evidence of the statements above
based on explicit computations, but a rigorous mathematical proof based on a geometric construction of the action of Virasoro or W-algebra is yet to be achieved.

\section{Instanton partition function with surface operator}\label{with}
\subsection{Surface operator}\label{surface-def}
A surface operator  creates a singularity of the gauge potential.
Surface operators for $\cN=4$ super Yang-Mills theory were studied by Gukov and Witten \cite{Gukov:2006jk}.
The $\cN=2$ version has been less studied, see e.g.~\cite{Alday:2009fs,Kozcaz:2010af,Gukov:2007ck,Tan:2009qq,Gaiotto:2009fs}.
Mathematically, surface operators in $\cN=2$ theories were first explored by Kronheimer and Mrowka \cite{KronheimerMrowka1,KronheimerMrowka2}.

Consider a $\U(N)$ gauge field on $\bR^4\simeq \bC^2$ parameterized by two complex variables $(z_1,z_2)$.
Let us put the surface operator at $z_2=0$, filling the $z_1$-plane.
We introduce polar coordinates on the $z_2$-plane, namely $z_2=r\exp(i\theta)$.
Then the surface operator creates the singularity \begin{equation}
A_\mu dx^\mu \sim \diag(\alpha_1,\ldots,\alpha_N) id\theta
\label{polar}
\end{equation} close to $r\sim 0$.
By a suitable gauge transformation we can assume $0\le \alpha_i <1$
and $\alpha_i \le \alpha_{i+1}$.
We consider the most generic case where the subgroup which commutes with the divergent part \eqref{polar} is $\U(1)^N$,
i.e.~the $\alpha_i$ are all distinct.
We call this a \emph{full} surface operator.\footnote{The surface operator which corresponds to the insertion of degenerate fields the Liouville/Toda theory \cite{Alday:2009fs,Drukker:2010jp} has $(\alpha_1,\ldots,\alpha_N)\propto (N-1,-1,\ldots,-1)$ .
The commuting subgroup is then $\U(1)\times \U(N-1)$,
and can be called the \emph{simple} surface operator.
For $N=2$ there is no difference between a full surface operator and a simple surface operator; we will come back to this point in Sec.~\ref{conclusions}. } 

The gauge group  on the surface operator  at $z_2=0$ is restricted to $\U(1)^N$, in order for it to be compatible with the singularity.
Correspondingly, there are $N$ Abelian gauge fields $F_{1}$,\ldots, $F_N$ or equivalently line bundles $L_1$,\ldots, $L_N$ on the surface operator.
To a given gauge configuration, we can associate the monopole numbers \begin{equation}
\ell_i=\frac1{2\pi}\int_{z_2=0} F_i =\int_{z_2=0} c_1(L_i).
\end{equation}
We only want $\U(N)$ configurations whose non-trivial curvature is in $\SU(N)$.
Therefore we require $\sum \ell_i=0$.

To define the instanton number, we define a smooth configuration $\bar A_\mu$ on $\bR^4$ by \begin{equation}
A_\mu dx^\mu = \bar A_\mu dx^\mu + f(r)  \diag(\alpha_1,\ldots,\alpha_N) id\theta
\end{equation} where $f(r)$ is a smooth function which satisfies $f(0)=1$ and $\lim_{r\to \infty} f(r)=0$.
Then we \emph{define} the instanton number $k$, which is an integer, of the original configuration $A_\mu$ to be the instanton number of $\bar A_\mu$.
Note that this does not equal the integral of $\tr F_A\wedge F_A$.
Instead, a short calculation shows that \begin{equation}
\frac{1}{8\pi^2} \int_{\bC^2\setminus\{z_2=0\}} \tr F_A\wedge F_A= k + \frac12\sum\alpha_i \ell_i.
\end{equation}
It means, in particular, that when the $\theta$ term is included in the Lagrangian, not only the instanton number $k$ but also the monopole numbers $\ell_i$ and weights $\alpha_i$ naturally enter in the path integral.

\subsection{Moduli space of instantons with surface operator}\label{surface-moduli}
We would like to consider the moduli space of anti-self-dual
gauge configurations with the singularity \eqref{polar} at the
surface operator. As recalled in the previous section, the
topological data to be specified are $k$ and $\ell_i$. It turns
out to be more convenient to use $\vec k=(k_1,\ldots,k_N)$
defined by the relations \begin{equation} k_1=k,\qquad
k_{i+1}=k_i+\ell_i.
\end{equation}
We denote the moduli space of such configurations by $\cM_{N;\vec k}$.
It is known to be a space of real dimension $4(k_1+\cdots+k_N)$, but again with various singularities.
We would like to consider the moduli space of slightly different objects in order to cure these singularities.
Without the surface operator, we could use the non-commutative deformation.
Presumably this is still possible with the surface operator, but the study of surface operators in the non-commutative plane is not yet available.
We can instead use the language of sheaves.
Here, we use a theorem conjectured in \cite{KronheimerMrowka1,KronheimerMrowka2} and proved in \cite{Biquard}, which establishes a one-to-one mapping between
\begin{itemize}
\item Anti-self-dual gauge configurations on $\bR^4$ which are smooth away from $z_2=0$ and with the behavior \eqref{polar}.
\item Stable rank-$N$ locally-free sheaves on $\CP^1\times \CP^1$
with a parabolic structure at $\{z_2=0\}$
and with a framing at infinities, $\{z_1=\infty\}\cup\{z_2=\infty\}$.
\end{itemize}
The moduli space of the latter can be made into a smooth space by considering not just locally-free sheaves but also torsion-free sheaves.
We denote this space by $\tilde\cM_{N,\vec k}$.
This is non-empty if and only if $k_i\ge 0$ for all $i$.
Then it is a smooth complex manifold of complex dimension $2(k_1+\ldots+k_N)$, and sometimes called the affine Laumon space in the mathematics literature.

The spacetime rotation $\SO(4)$ is broken down to $\U(1)\times \U(1)$ because of the replacement of $\bR^4$ by $\CP^1\times \CP^1$.
The global gauge rotation $\U(N)$ still acts on these configurations.
Then we can consider the fixed points of the action of $\U(1)^2\times \U(1)^N\subset \U(1)^2\times \U(N)$ on $\tilde \cM_{N,\vec k}$.
These were enumerated in \cite{FFNR}.
They are still labeled by an $N$-tuple of Young diagrams
$\vec \eth=(\eth_1,\ldots,\eth_N)$ but with the following constraints.
Let $\eth_i = (\eth_{i,1}\ge \eth_{i,2}\ge \cdots)$ be the $i$-th Young diagram.
We then define $\eth_{i,j}$ for $i$ outside the range $1$,\ldots,$N$ by requiring the condition $\eth_{i,j}=\eth_{i+N,j}$.
We then define \begin{equation}
k_i(\vec\eth)=\sum_{j\ge 0} \eth_{i+j,j+1}.
\end{equation}
Then the fixed points of $\cM_{N,\vec k}$ are labeled by  vectors of Young diagrams $\vec\eth$ such that  $k_i=k_i(\vec\eth)$.

\subsection{Instanton partition function}

We define the instanton partition function in the presence of the surface operator by the same expressions \eqref{integral}, \eqref{sum} with $\tilde\cM_{N,k}$  replaced by $\tilde \cM_{N,\vec k}$. We believe this coincides with the partition function in the omega background in the physical sense, but this point deserves to be further clarified.

First let us consider $\cN=2$ $\U(N)$ theory with an adjoint hypermultiplet of mass $m$.
Then the instanton partition function in the presence of the full surface operator  is \begin{equation}
Z^S(x_1,\ldots,x_N ; \vec a ; m ) =
\sum_{\vec\eth} x_1^{k_1(\vec\eth)}\cdots x_N^{k_N(\vec\eth)}
\frac{z^S_\text{adj}[\vec\eth](\vec a,m)}{z^S_\text{vector}[\vec\eth](\vec a)}.
\label{partition-with-surface}
\end{equation}
Here, $x_i$ is the counting parameters
for the topological number $k_i$ of the instantons;
$z^S_\text{vec}[\vec\eth](\vec a)$ and $z^S_\text{adj}[\vec\eth](\vec a)$ are the contribution of the vector multiplet and of the adjoint hypermultiplet
when the gauge configuration is $\vec\eth$ and the vacuum expectation value is $\vec a$.

$z^S_\text{vec}$ and $z^S_\text{adj}$ can be written easily
using the contribution of a bifundamental multiplet
$z^S_\text{bif}[\vec\eth,\vec\eth'](\vec a,\vec a',m)$ coupled
to the gauge configurations $\vec\eth$, $\vec\eth'$ and the
vacuum expectation values $\vec a$, $\vec a'$ of the first and
the second gauge groups \begin{equation}
z^S_\text{vec}[\vec\eth](\vec
a)=z^S_\text{bif}[\vec\eth,\vec\eth](\vec a,\vec a;0),\qquad
z^S_\text{adj}[\vec\eth](\vec
a,m)=z^S_\text{bif}[\vec\eth,\vec\eth](\vec a,\vec a;m).
\end{equation}

The bifundamental contribution itself can be computed by the
following procedure. 
To given gauge configurations $\vec\eth$, $\vec\eth'$ of the two gauge groups, 
we associate a vector space of complex dimension $|\vec\eth|+|\vec\eth'|$, 
which we  denote by $\Ext(\vec\eth,\vec\eth')$.
This would be
the space of zero-modes of the Dirac operator in the
bifundamental representation coupled to the gauge
configurations $\vec\eth$ and $\vec\eth'$  if these
configurations were genuine smooth gauge fields.

Spacetime rotations $\U(1)^2$
and  gauge rotations $\U(N)\times \U(N)$ naturally act on this vector space.
Take a typical group element $g$ in this symmetry group
\begin{equation}
g=\exp(\diag\{\epsilon_1,\epsilon_2;a_1,\ldots,a_N;a'_1,\ldots,a'_N\}).
\end{equation} Then the trace $\tr_{\Ext({\vec\eth},{\vec\eth'})} g$ is given by \cite{FFNR}
\begin{align}
\tr_{\Ext({\vec\eth},{\vec\eth'})} g = &\sum_{k=1}^N\sum_{\substack{l \le k \\ l'\le k-1}}
\frac{e^{a_l}}{e^{a_{l'}'}}
e^{\epsilon_1}e^{\epsilon_2(\lfloor\frac{-l'}{N}\rfloor-\lfloor\frac{-l}{N}\rfloor)} \frac{(e^{\epsilon_1d'_{k-1,l'}}-1)(e^{-\epsilon_1d_{k,l}}-1)}{e^{\epsilon_1}-1} \nonumber\\
&+
\sum_{k=1}^N\sum_{l'\le k-1}
\frac{e^{a_k}}{e^{a_{l'}'}}
e^{\epsilon_1}e^{\epsilon_2(\lfloor\frac{-l'}{N}\rfloor-\lfloor\frac{-k}{N}\rfloor)} \frac{e^{\epsilon_1d'_{k-1,l'}}-1}{e^{\epsilon_1}-1} \nonumber\\
&-\sum_{k=1}^N\sum_{\substack{l \le k \\ l'\le k}}
\frac{e^{a_l}}{e^{a_{l'}'}}
e^{\epsilon_1}e^{\epsilon_2(\lfloor\frac{-l'}{N}\rfloor-\lfloor\frac{-l}{N}\rfloor)} \frac{(e^{\epsilon_1d'_{k,l'}}-1)(e^{-\epsilon_1d_{k,l}}-1)}{e^{\epsilon_1}-1} \nonumber\\
&-
\sum_{k=1}^N\sum_{l\le k}
\frac{e^{a_l}}{e^{a_k'}}
e^{\epsilon_1}e^{\epsilon_2(\lfloor\frac{-k}{N}\rfloor-\lfloor\frac{-l}{N}\rfloor)} \frac{e^{-\epsilon_1d_{k,l}}-1}{e^{\epsilon_1}-1}.\label{character}
\end{align}
Here, $d_{ij}$  is defined by $d_{ij}=\eth_{j,i-j+1}$
and $a_{l}$ for $l$ outside the range $1$,\ldots, $N$  is defined so that $a_{l+N}=a_l$.

Finally the contribution of a bifundamental is given by \begin{equation}
z^S_\text{bif}[\vec\eth,\vec\eth'](\vec a,\vec a';m)=\prod_{i=1}^{|\vec\eth|+|\vec\eth'|} (s_i-m)
\end{equation} where $s_i$ arise in the expansion of the trace  in monomials
\begin{equation} \tr_{\Ext(\vec\eth,\vec\eth')} g =
\sum_{i=1}^{|\vec\eth|+|\vec\eth'|} e^{s_i}.
\end{equation}  $s_i$ is a linear combination of $\epsilon_{1,2}$, $a_{1,\ldots,N}$, and $a_{1,\ldots,N}'$.
As the calculation is quite contrived, we provide a few explicit examples in Appendix~\ref{examples}.

\subsection{Reformulation}
Now our strategy is the same as in the case without surface
operator, reviewed in Sec.~\ref{reformulation}. We define the
vector space $\cH^S_{\vec a}$ whose basis
$\kket{\vec\eth}_{\vec a}$ is labeled  by  the fixed points in
$\tilde \cM_{N,\vec k}$, and introduce the following inner
product in such space
\begin{equation} {}_{\vec a}\vvevv{\vec\eth|\vec\eth'}_{\vec
a}=\delta_{\vec\eth,\vec\eth'}/z^S_\text{vector}[\vec\eth](\vec
a).
\end{equation}
We also introduce operators $\bk_i$ which measure the instanton numbers: \begin{equation}
\bk_i \kket{\vec\eth} \equiv k_i(\vec\eth)\kket{\vec\eth}.
\end{equation}
Finally we define the operators $\Phi^S_{\vec a,m,\vec b}:\cH^S_{\vec b}\to \cH^S_{\vec a}$ via \begin{equation}
\Phi^S_{\vec a,m,\vec b}\kket{\vec\eth}_{\vec b}=\sum_{\vec \eth'} \frac{z^S_\text{bif}[\vec \eth',\vec\eth](\vec a,\vec b,m)}{z^S_\text{vector}[\vec\eth](\vec b)}\kket{\vec \eth'}_{\vec a}.
\end{equation}

Then the partition function with the surface operator \eqref{partition-with-surface} can be recast into the form \begin{equation}
Z^S(x_1,\ldots,x_N;\vec a;m)=\tr_{\cH^S_{\vec a}} \Phi^S_{\vec a,m,\vec a} x_1^{\bk_1}\cdots x_N^{\bk_N}.
\end{equation}

Here comes a surprise: it is a \emph{proven mathematical theorem} \cite{Braverman:2004vv,Braverman:2004cr,FFNR} that $\cH^S_{\vec a}$ has a natural geometric action of the affine $\SL(N)$ algebra. The proven properties are the following:
\begin{enumerate}
\item The level is given by $k=-N-\epsilon_2/\epsilon_1$.
\item Let us denote by $\bV_{\vec\jmath}$ the Verma module of the affine $\SL(N)$ algebra associated to a primary state of spin $\vec\jmath$.
Then $\cH^S_{\vec a}$ splits as\footnote{Strictly speaking, it is only proved that there is a natural action of affine $SL(N)$ on $\cH^S$. This decomposition is conjectured in \cite{FFNR}. }  \begin{equation}
\cH^S_{\vec a}=V_\text{Fock}\otimes \bV_{\vec a/\epsilon_1-\vec\rho}.
\end{equation} where $V_\text{Fock}$ is the Fock space of a free boson.
\item The operators $\bk_i$ have a natural action on $\bV_{\vec\jmath}$\ . \ 
For instance, $\sum_i \bk_i$ measures $L_0$ of a descendant with respect to the primary, 
whereas $\bk_i$ measures $H_i$ of a descendant with respect 
to  the primary, where $H_{1},$ \ldots $H_{N-1}$ denote the standard Cartan generators associated to the nodes of the Dynkin diagram of the non-affine $SL(N)$ algebra.
For example for $N=2$, $\bk_1=2J^0_0$ and $\bk_2=L_0-2J^0_0$.
\end{enumerate}
The next subsection will be devoted to the study of the operator $\Phi^S_{\vec a',m,\vec a}$.

\subsection{Partition function as conformal blocks}\label{block}
We would like to understand the linear operator $\Phi^S_{\vec a',m,\vec a}$ acting on the space $\cH^S_{\vec a}$.
$\cH^S_{\vec a}$ is a tensor product of a Fock space and the Verma module of the affine Lie algebra.
Therefore we expect the operator $\Phi^S$ to decompose into the product of the free boson vertex operator times the insertion of a primary\footnote{A case which is mathematically slightly simpler was studied by \cite{Negut}, where the geometric realization of the non-affine $\SL(N)$ was considered. There, it was noticed that there is a small modification to the  insertion of a primary operator,
and therefore we also expect that there is a similar small modification in our case.
}, as was the case in \cite{Alday:2009aq}.
The precise form can be determined experimentally by calculating the instanton partition function with the surface operator for various $\cN=2$ gauge theories.

We performed an extensive calculation in the case $N=2$,
and obtained the following results.
For the $\SU(2)$ theory with an adjoint hypermultiplet,
the instanton partition function with the surface operator is
\begin{equation}
Z=\tr_{\cH^S_{\vec a}} \Phi^S_{\vec a, m ,\vec a} x_1^{\bk_1}x_2^{\bk_2}.
\end{equation} We found that it equals \begin{equation}
Z=\left[\prod_{i=1}^\infty (1-z^i)^{m(2\epsilon_1+\epsilon_2-2m)/(\epsilon_1\epsilon_2) -1 } \right] \tr_{V_j} \cK \bV_{-m/\epsilon_1}(x,1) z^{\bf N}.
\end{equation} Here, $\bV_j(x,z)$ is the primary field of spin $j$
as described in Appendix~\ref{appendix}; we take $x=x_1$ and $z=x_1x_2$.
The operator $\cK$ is given by the exponential of raising operators \begin{equation}
\cK=\exp(- \sum_{i=1}^\infty \frac{1}{2i-1}[J^-_{1-i} + J^+_{-i}] ).
\end{equation}

Similarly, the partition function for $\SU(2)$ theory with four flavors is given by \begin{equation}
Z={}_{a_1}\bbra{\vec\varnothing} \Phi^S_{a_1,m_1,a}
x^{\bk_1} y^{\bk_2} \Phi^S_{a,m_2,a_2} \kket{\vec\varnothing}_{a_2}.
\end{equation}
Here, as in our previous paper \cite{Alday:2009aq}, 
the mass parameters $a_{1,2}$ and $m_{1,2}$ are given in terms of
the four hypermultiplet masses $M_{1,2,3,4}$ by \begin{equation}
a_1=M_1-M_2,\quad
m_1=M_1+M_2,\quad
a_2=M_3-M_4,\quad
m_2=M_3+M_4.
\end{equation}
We found the partition function to be equal to \begin{equation}
Z=(1-z)^{m_1(2\epsilon_1+\epsilon_2-2m_2)/(\epsilon_1\epsilon_2)}
\vev{j_1|\bV_{j_2}(1) \cK  \bV_{j_3}(x,z)|j_4}
\end{equation} where \begin{equation}
j_1=\frac{a_1}{\epsilon_1}-\frac12,\quad
j_2=-\frac{m_1}{\epsilon_1},\quad
j=\frac{a}{\epsilon_1}-\frac12,\quad
j_3=-\frac{m_2}{\epsilon_1},\quad
j_4=\frac{a_2}{\epsilon_1}-\frac12.
\end{equation} Here $j$ is the spin chosen to be in the intermediate channel. 

Hence, the partition function in the presence of the full surface operator
coincides with the four-point conformal block on the sphere of affine $SL(2)$  Lie algebra,
with the additional insertion of $\cK$. 
In the same way, one can obtain the five-point conformal block on the sphere
as well as the one- and two-point conformal blocks on the torus with the  insertion of the same operator $\cK$.

From these computations, we find that the operator $\Phi_{a',m,a}$
has a decomposition \begin{equation}
\Phi_{a',m,a}= V_{2\epsilon_1+\epsilon_2-2m,m} \otimes \cK\bV_{-m/\epsilon_1} :
V_\text{Fock} \otimes \bV_{j(a)} \to
V_\text{Fock} \otimes \bV_{j(a')},
\end{equation} where $j(a)=a/\epsilon_1-1/2$.

\section{Partition functions as eigenfunctions}\label{eigen}
\subsection{$\cN=2$ theory and the quantized integrable systems}
To an $\cN=2$ theory in four dimensions,
one can naturally associate a classical  holomorphic integrable system \cite{Martinec:1995by,Donagi:1995cf}.
When the $\cN=2$ theory is obtained by wrapping $N$ M5-branes on a punctured Riemann surface $C_2$ as in \cite{Gaiotto:2009we}, the holomorphic integrable system associated to it is the $\SU(N)$ Hitchin system associated to $C_2$.

For example, consider wrapping $N$ M5-branes on a torus with a
simple puncture. The resulting $\cN=2$ theory is the $\SU(N)$
theory with a massive adjoint hypermultiplet, which is often called the $\cN=2^*$ theory \cite{Donagi:1995cf,Itoyama:1995nv}.
The Coulomb
branch is parameterized by the vacuum expectation values $\tr
\phi^k$ for $k=2,\ldots,N$, where $\phi$ is the scalar
component of the $\SU(N)$ vector multiplet. Then the integrable
system associated to it is the elliptic Calogero-Moser model of
$N$ particles on $T^2$; here $\tr\phi^2$ corresponds to the
usual Hamiltonian and $\tr\phi^k$ with $k>2$ give other
commuting Hamiltonians. Moreover, the Lax matrix of the
elliptic Calogero-Moser model can be  naturally viewed as a
field $\Phi(z)$ in the adjoint of $\SL(N)$ on $T^2$ with a
given singularity, where the coordinate of the torus serves as
the spectral parameter  \cite{Gorsky:1994dj}. Indeed it is a
generic feature of all integrable models obtained from the
Hitchin system on a punctured Riemann surface $C_2$ that the
Higgs field $\Phi(z)$ on $C_2$ serves as the Lax matrix.

\begin{figure}
\[
\includegraphics[width=.3\textwidth]{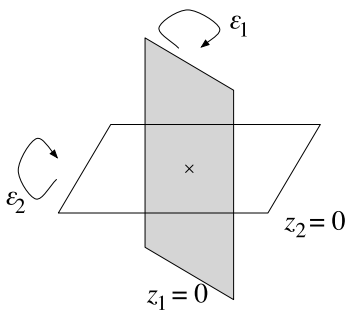}
\]
\caption{The setup of Nekrasov-Shatashvili and the addition of a surface operator. We insert the surface operator on the plane $z_2=0$, which is shown as an unshaded parallelogram. In the limit $\epsilon_2\to 0$,
we have an effectively two-dimensional system on the plane $z_1=0$,
shown as an shaded parallelogram.
\label{configuration}}
\end{figure}

Recently Nekrasov and Shatashvili made a striking observation
\cite{Nekrasov:2009rc} concerning this integrable system
associated to the gauge theory, which we now briefly recall. As
reviewed above, the deformations by $\epsilon_1$ and
$\epsilon_2$ effectively confine the excitations close to the
origin of the spacetime. Now consider the limit
$\epsilon_2\to 0$, keeping $\epsilon_1$ finite. The excitations
are then confined to the plane $z_1=0$ but they are free to
move in the $z_2$ direction, see figure~\ref{configuration}.
The system then behaves as a two-dimensional theory. The
preserved supersymmetry  turns out to be $\cN=(2,2)$, and
therefore the dynamics of the vector multiplet is governed by
the twisted superpotential $\cW(\epsilon_1,\vec a,\vec m)$.
Analogously to \eqref{limit-to-prepotential}, the twisted
superpotential can be recovered from the partition function 
in the limit $\epsilon_2\to 0$
\begin{equation} Z(\epsilon_1,\epsilon_2;\vec a,\vec m) =
\exp (-\frac{\cW(\epsilon_1,\vec a,\vec m)}{\epsilon_2}+\cdots
).
\end{equation} 
The vacuum expectation values of $\vec a$ are determined by the conditions \begin{equation}
\exp(\frac{\partial\cW}{\partial a_i})=1.\label{Bethe}
\end{equation}
The observation was that this set of equations is the Bethe
ansatz equations of a quantum integrable model, and that this
quantum model is a quantized version of the integrable system
associated to the $\cN=2$ gauge theory.

To obtain the vacuum expectation values of gauge invariant operators,
we first perturb the ultraviolet action \begin{equation}
\int d^4\theta \tau_{UV} \tr \phi^2
\to
\int d^4\theta (\tau_{UV} \tr \phi^2  + t_k \tr\phi^k).
\end{equation}
Second, we  calculate the partition function $Z$ and the
twisted superpotential $\cW$ in the presence of these
perturbations. The vacuum expectation values of gauge
invariant operators are then given by
\begin{equation} \vev{\tr\phi^k} =
\frac{\partial\cW(\epsilon_1,\vec a,\vec m,\vec t)}{\partial
t_k}\Big|_{t_k=0}
\end{equation} evaluated at solutions of \eqref{Bethe}.
For $k=2$ the situation is particularly simple, as we don't
need to introduce $t_2$ :
\begin{equation}
\vev{\tr\phi^2} = \frac{\partial\cW(\epsilon_1,\vec a,\vec m)}{\partial \tau_{UV}}.\label{u}
\end{equation}
The quantization of the classical integrable system involves
two steps. The first is to obtain commuting differential
operators whose classical limit reproduce the commuting
Hamiltonians of the integrable system. At this stage we only
have operators and we do not have states to act on. The problem
is particularly severe in this case, because the original
integrable system was holomorphic and thus the differential
operators are all holomorphic objects. The second
step is to choose a particular real section of the space and
supply a Hilbert space for the operators to act on. We do not
have much to comment on this second step, but as we will soon
see the introduction of the surface operator is  relevant
for the first step.

\subsection{Conformal blocks and the quantized Hitchin system}

We saw in Sec.~\ref{surface-moduli} that the instanton moduli space with the surface operator naturally has the action of affine $\SL(N)$ algebra with level $k=-N-\epsilon_2/\epsilon_1$.
In the limit $\epsilon_2\to 0$ which Nekrasov and Shatashvili considered, the level becomes \emph{critical} and the Sugawara construction fails.
Recall that the energy-momentum tensor $T(z)$ is given by \begin{equation}
T(z)=\frac{2}{k+N} S(z)\quad
\text{where}
\quad S(z)={: J^a(z)J^a(z):}\ ,
\end{equation}  and has the OPE \begin{equation}
T(z)T(0)\sim \frac{c}{2z^4} + \frac{2T(0)}{z^2} +\frac{\partial T(0)}{z}
\quad\text{where}
\quad c=\frac{k \dim\SU(N)}{k+N}.
\end{equation}
Therefore, $T(z)$ becomes ill-defined when $k=-N$.
However, the operator $S(z)$ is still well-defined, and
moreover its OPE with itself and with the current $J^a$
is completely regular: \begin{equation}
S(z)S(0)\sim \cO(1),\qquad
S(z)J^a(0)\sim \cO(1).
\end{equation}
In other words, $S(z)$ essentially behaves as a classical
quadratic differential and the modes
of $S(z)$ commute with $J^a(z)$.

This property can be used to obtain the differential operators
which quantize the Hamiltonians of the Hitchin system on the
punctured Riemann surface \cite{BeilinsonDrinfeld}. Here we
will briefly recall how this construction goes. For reviews,
see e.g.~ Part III of \cite{Frenkel:2005pa}.

Let us recall the derivation of the Knizhnik-Zamolodchikov equations \cite{Knizhnik:1984nr}.
Consider the following conformal block  \begin{equation}
\vev{\cO_{j_1}(z_1)\cO_{j_2}(z_2)\cdots \cO_{j_n}(z_n)}.
\end{equation}
Its derivative with respect to $z_i$ can be given as a suitable contour integral of \begin{equation}
\vev{T(z)\cO_{j_1}(z_1)\cO_{j_2}(z_2)\cdots \cO_{j_n}(z_n)}
\label{TOO}
\end{equation}
in the plane parametrized by $z$. Next we rewrite $T(z)$ in
terms of the Sugawara construction. Then we use the OPE of
$J^a(z)$ with $\cO_j$ to express \eqref{TOO} in terms of the
action of the spin-$j$ representation on the extra label
carried by $\cO_j$. When the spin $j$ does not give a finite
dimensional representation, the action is typically given by
differential operators, as in Appendix~\ref{appendix}. Then we
have a relation of the form
\begin{equation} \frac{\partial}{\partial
z_i}\vev{\cO_{j_1}(z_1)\cO_{j_2}(z_2)\cdots \cO_{j_n}(z_n)}  =
\cD_{(i)} \vev{\cO_{j_1}(z_1)\cO_{j_2}(z_2)\cdots
\cO_{j_n}(z_n)} \label{KZ}
\end{equation} where $\cD_{(i)}$ is a certain second-order differential operator.

The derivative of the conformal block with respect to the
moduli of the Riemann surface is also given by an 
integral of \eqref{TOO} on a suitable contour of $z$, and
results in an equation similar to \eqref{KZ} where
$\partial/\partial z_i$ is replaced by $\partial/\partial \tau$
and $\cD_{(i)}$ is replaced by another certain differential
operator $\cD_{(\tau)}$. This equation is called the
Knizhnik-Zamolodchikov-Bernard equation \cite{Bernard:1987df}.
A notable fact is that the differential operator $\cD_{(\tau)}$ for the torus one-point function
is the quantized version of the elliptic Calogero-Moser Hamiltonian \cite{Etingof:1993gk,Felder:1994gk}\footnote{Martinec pondered on the relation of this fact with the gauge theory long ago \cite{Martinec:1995qn}. Relation to the Hitchin system was also mentioned in this connection in \cite{Bonelli:2009zp}.}.

So far we assumed $k\ne -N$. One can also perform this
procedure when the level is critical by inserting $S(z)$
instead of $T(z)$. The conformal block with an insertion of
$S(z)$ can still be rewritten as the conformal block acted by a
second-order differential operator. But at the critical level,
$S(z)$ does not have the power to change the moduli. Instead,
it just behaves as a c-number. In this way, one obtains a set
of second-order differential operators whose eigenvalues are
specified by a quadratic differential on the Riemann surface.

Furthermore, when $k=-N$,
it has been shown that  the operators \begin{equation}
S_k(z) = {:\tr (J^a(z)t_a)^k :}
\end{equation} have no singular terms in the OPE among themselves and with $J^a(z)$, just as $S(z)=S_2(z)$ did.
Insertion of $S_k(z)$ into the correlator then gives
a set of order-$k$ differential operators whose eigenvalues are given by a degree-$k$ differential on the Riemann surface.
These differential operators commute among themselves, because the OPEs among $S_k(z)$ are regular.
These are the differential operators which quantize the commuting Hamiltonians of the Hitchin system \cite{BeilinsonDrinfeld},
and they naturally act on the conformal blocks.

Note that in \cite{Nekrasov:2009rc} the gauge theory in the limit $\epsilon_2\to0$ provided the vacuum expectation values of $\tr \phi^k$, i.e.~the eigen\emph{values} of the quantized Hamiltonians.
Here the conformal blocks provide the eigen\emph{functions} of the quantized Hamiltonians.
Had we \emph{not} have the insertion of the extra operator $\cK$ inserted in the conformal block corresponding to the surface operator, we could directly use these beautiful mathematical results to conclude that
the partition function of the surface operator in the limit $\epsilon_2\to 0$ gives the simultaneous eigenfunction of the quantized Hamiltonians of the Hitchin system.

We need to study how the presence of $\cK$ modifies the
derivation. Note that $\cK$ is a particular element of the
affine Lie \emph{group}. $\cK$ acts by conjugating $J^a_n$ when
one moves the insertion of $J^a_n$ inside the correlation
functions in the derivation of the KZ equation, and $J^a_n$
after conjugation still acts on the primary field by a linear
differential operator. Therefore, $S_k(z)$ still gives an
order-$k$ differential operator. We expect that they just give
a different quantization of the Hitchin Hamiltonians. This
point clearly needs to be investigated in more detail in the
general case. Instead, in the following we consider as a
concrete example the $\cN=2^*$ $\SU(2)$ theory, where the
insertion of $\cK$ produces very mild effects, and study how
this works in detail.

\subsection{Torus one-point function and Calogero-Moser Hamiltonian}

The integrable system associated to the $\cN=2^*$ theory
is the Hitchin system on the torus with one puncture \cite{Donagi:1995cf}.
This is equivalent to the elliptic Calogero-Moser system \cite{Gorsky:1994dj}.
Let us first consider the ordinary conformal block of affine $\SL(2)$ algebra without the insertion of $\cK$: \begin{equation}
F(x,z)= x^{-j_i} z^{j_i(j_i+1)/(k+2)}\tr_{\bV_{j_i}} z^{\bf N} \bV_{j_e}(x,1) .
\end{equation} The KZB equation is then given by \cite{Etingof:1993gk} \begin{equation}
(k+2) z\frac{\partial}{\partial z} \cF(x,z)
= \cH_x \cF(x,z)  \label{KZB}
\end{equation}
where
\begin{equation}
\cH_x= x\frac{\partial}{\partial x}(x\frac{\partial}{\partial x}-1)
+j_e(j_e+1) \cP(x,z)
.
\end{equation}
Here, $\cF(x,z)$ is the normalized one-point function,
\begin{equation} \cF(x,z)=
\frac{F(x,z)}{F(x,z)|_{j_e=0}}=(1-x-z/x) F(x,z)
\end{equation} and $\cP$ is the Weierstrass elliptic function
given by
\begin{eqnarray}
\cP(x,q)&=&(x \partial_x)^2 \log \theta(x,q), \quad \text{where}\\
\theta(x,q)&=&i q^{1/8} (x^{1/2}-x^{-1/2})\prod_{n=1}^\infty (1-q^n x)(1-q^n/x)(1-q^n).
\end{eqnarray}
$\cH_x$ is the elliptic Calogero-Moser Hamiltonian for the two-particle system.

We now consider the conformal block with the insertion of $\cK$:
\begin{equation}
F_{\cK}(x,z)= x^{-j_i} z^{j_i(j_i+1)/(k+2)}\tr_{\bV_{j_i}} z^{\bf N} \cK\bV_{j_e}(x,1) .
\end{equation}
Explicit calculation shows that it satisfies\footnote{This
equality was checked to order cubic in $x$ and $z/x$ by
Mathematica. For a closely related calculation for non-affine
$SL(n)$ algebra, this relation was proved in \cite{Negut}. It
should  also be noted that this simple relation between
conformal blocks with and without $\cK$ does not easily extend
to other cases, e.g.~the two-point function on the torus.}
\begin{equation}
F_{\cK}(x,z)= (1-x-z/x)^{-j_e}F(x,z).
\end{equation}
Therefore, $F_{\cK}$ satisfies the same equation as \eqref{KZB}
with the differential operators conjugated by $(1-x-z/x)^{j_e}$.
This can be thought of as a normal-ordering ambiguity.

Let us take  the critical limit $k \rightarrow -2$. Naively,
the derivative with respect to $z$ on the left hand side of
\eqref{KZB} drops out. However this limit is subtle, since the
perturbative expansion computing the conformal block has poles
at $k=-2$ which need to be taken care of. It can be seen that
such poles exponentiate nicely  and we have the following
structure\footnote{The four-point conformal block on the sphere is expected to have this behavior \cite{Ribault:2005wp,Reshetikhin:1994qw}. }
\begin{equation}
\label{expon}
\cF(x,z)= z^{j_i(j_i+1)/(k+2)}
e^{W(z)/(k+2)} \cF_\text{reg}(x,z)
\end{equation}
where
$\cF_\text{reg}(x,z)$ is perfectly finite when $k\to -2$.
Here, we keep $j_i$ and $j_e$ fixed when taking the limit.
It is very important that $W(z)$ does not depend on $x$.
We have checked this up to five instanton order.

Let $\cF_\text{reg}^{(0)}=\lim_{k\to -2} \cF_\text{reg}$.
Plugging (\ref{expon}) into \eqref{KZB} and taking the limit,
we learn that
\begin{equation}
{\cal H}_x {\cal F}_\text{reg}^{(0)}
= \left[ j_i(j_i+1) + z W'(z) \right] {\cal F}_\text{reg}^{(0)}.\label{value}
\end{equation}
Therefore, ${\cal F}_\text{reg}^{(0)}$ is an eigenfunction of the
Calogero-Moser Hamiltonian. Note that from this point of view
$z$ is a parameter, not associated to a degree of freedom.

Let us study the Virasoro conformal block in the same limit.
The torus one-point function is \begin{equation}
Z=\tr_{\cV_{\alpha_i}} z^{\bf N}\cV_{\alpha_e}(1).
\end{equation}
One finds that it behaves as
\begin{equation}
Z=e^{(\epsilon_1/\epsilon_2)\cW_\text{inst}(z)} Z_\text{reg}(z)
\end{equation} in the limit $\epsilon_2\to 0$, with
 $Z_\text{reg}(z)$ being finite.
Moreover, we find\footnote{Again, this equality was checked via
explicit calculations. For the pure $\SU(N)$, the corresponding
statement was proved in \cite{Braverman:2004vv}.}
\begin{equation} \cW_\text{inst}(z,\alpha_i,\alpha_e) = W(-z,j_i,j_e)
\end{equation} with the following identifications, 
\begin{align}
\alpha_i= -b j_i + b^{-1}/2, \quad
\alpha_e = -b j_e
\end{align}  where $b=\sqrt{\epsilon_1/\epsilon_2}$.
This is chosen so that they correspond to the same gauge theory parameters.
We can give the following gauge theory interpretation to what we have just found.
In the $\epsilon_2\to 0$ limit, the system decompactifies to a theory on the $z_1=0$ plane, see figure~\ref{configuration}.
The factor $1/(k+2)=\epsilon_1/\epsilon_2=b^2$ measures the effective area of this $z_1=0$ plane, and both $W(z)$ and $\cW(z)$ give the bulk contribution to the partition function.
The surface operator is at $z_2=0$ and thus its contribution is independent of the effective area. Therefore it is quite natural that $W(z)=\cW(z)$, and indeed $\cW(z)$ is precisely the instanton contribution to the twisted superpotential.
Using the formula \eqref{u} expressing $\vev{\tr\phi^2}$ in the gauge theory in terms of  the twisted superpotential, we find that the relation \eqref{value} can be rewritten as \begin{equation}
{\cal H}_x {\cal F}_\text{reg}^{(0)}
= (j_i(j_i+1)+\vev{\tr\phi^2}_\text{inst}) {\cal F}_\text{reg}^{(0)}.
\end{equation}
Note that $j_i(j_i+1)$ is the classical contribution to $\vev{\tr\phi^2}$.
Recall that the classical Hitchin Hamiltonians corresponded to the vev of the gauge invariant operators.
We found that the finite part of the $\epsilon_2\to 0$ limit of the partition function of the surface operator gives the eigenfunction of the quantized Hamiltonian, and the eigenvalue is the vev of the gauge invariant operator.

\section{Discussions and Conclusions}\label{conclusions}

In this paper we studied the instanton partition function of $\cN=2$ gauge theories obtained by wrapping $N$ M5-branes on a punctured Riemann surface in the presence of the full surface operator.
We found that the partition function is given by a modified conformal block of affine $\SL(N)$ algebra.
We also saw that in the limit $\epsilon_2\to 0$ the partition function of the surface operator in $\cN=2^*$ $\SU(2)$ theory gives the eigenfunction of the quantized Calogero-Moser Hamiltonian,
and conjectured that in general the limit of the partition function of the surface operator gives the simultaneous eigenfunction of quantized Hamiltonians of the Hitchin system associated to the Riemann surface with punctures.
Thus the partition function of the surface operator bridges the gauge-theoretical realization of  conformal field theory on the Riemann surface initiated in \cite{Alday:2009aq} and that of quantized Hitchin system in \cite{Nekrasov:2009rc}, in a way complementary to the work by \cite{Nekrasov:2010ka}.

In this paper we only scratched the surface of the problem.
The insertion $\cK$ for general $N$ should be determined.
The properties of conformal blocks with insertions of $\cK$ and the KZ equations they satisfy should be further studied.

Let us now summarize the mapping between the gauge theory and the CFT variables:
\begin{center}
\begin{tabular}{c|r@{}l|r@{}l}
$\SU(N)$ gauge theory &\multicolumn{2}{c|}{affine $\SL(N)$} & \multicolumn{2}{c}{$W_N$ algebra} \\
\hline
$\epsilon_{1,2}$ & $b$&$=\vphantom{\Bigm[}\sqrt{\epsilon_1/\epsilon_2}$ & $Q$&$=b+1/b$ \\[.5em]
&$k$&$=-N-b^{-2}$ & \multicolumn{2}{c}{$c=(N-1)(1+N(N+1)Q^2)$}\\[.5em]
hyper with mass $m$ &
spin $\vec\jmath(m)$&$\displaystyle=-\frac{Nm}{\epsilon_1}\vec\chi$ &
momenta $\vec\alpha(m)$&$\displaystyle=\frac{Nm}{\sqrt{\epsilon_1\epsilon_2}}\vec\chi $ \\[.6em]
vev $\vec a$ &
spin $ \vec \jmath(\vec a)$&$\displaystyle=\frac{\vec a}{\epsilon_1}-\vec\rho$ &
momenta $\vec\alpha(\vec a)$&$\displaystyle=\frac{\vec a+(\epsilon_1+\epsilon_2)\vec \rho/2}{\sqrt{\epsilon_1\epsilon_2}}$
\end{tabular}
\end{center}

One intriguing connection which we immediately see is the relation between the level and the central charge.
There is a method called the quantum Drinfeld-Sokolov reduction \cite{Bershadsky:1989mf,Feigin:1990pn} which extracts a representation of $W_N$ algebra given a representation of affine $\SL(N)$ algebra.
It is straightforward to confirm that the reduction of the affine algebra with $k=-N-b^{-2}$ gives the $W_N$ algebra with central charge $c=(N-1)(1+N(N+1)Q^2)$.

Under the reduction, the spin $\vec\jmath$ in the affine algebra and the momentum $\vec\alpha$ in the $W_N$ algebra are related by \begin{equation}
\vec\alpha=-b\vec\jmath,
\end{equation} see Appendix~\ref{free} for a short reminder.
Note that the relation between $\vec\jmath(m)$ and $\vec\alpha(m)$ is as expected from the  reduction.
However, the relation between $\vec\jmath(\vec a)$ and $\vec\alpha(\vec a)$ is not. It is instead  \begin{equation}
\vec\alpha=-b\vec\jmath + b^{-1}\vec\rho.\label{SRT-momenta}
\end{equation}
This is the mapping found in \cite{Ribault:2005wp} for $N=2$
and in \cite{Ribault:2008si} for $N=3$
when the correlation function of $SL(N)$ Wess-Zumino-Witten model
was converted to that of Liouville and Toda theory.

This observation reminds us of the special property of the surface operator when $N=2$.
As mentioned in Sec.~\ref{surface-def}, for $\SU(2)$ there is no difference between a simple surface operator and a full surface operator.
In this paper we argued that the insertion of a full surface operator changes the conformal field theory from the Virasoro algebra to the affine $\SL(2)$ algebra.
On the other hand, it was argued in \cite{Alday:2009fs,Drukker:2010jp} that the insertion of a simple surface operator corresponds to the  insertion of a certain degenerate field on the Riemann surface.
Note that the insertion of the degenerate field creates a surface operator for a particular $\SU(2)$ gauge group in the quiver gauge theory, while the change of the chiral algebra from the Virasoro to the affine $\SL(2)$ creates a surface operator for every $\SU(2)$ gauge group in the theory.
This reminds us of the mapping\footnote{The relation of this mapping in the context of the relation of the gauge theory to the instanton partition function  was also suggested in \cite{Giribet:2009hm}.}
of Ribault and Teschner between the correlation functions of $\SL(2)$ Wess-Zumino-Witten theory and of Liouville theory \cite{Ribault:2005wp,Hikida:2007tq,Giribet:2008ix}.
There, one needed to insert extra degenerate fields on the Riemann surface to establish the mapping.
This agrees at a rough level with our observation that we need to insert multiple degenerate fields to create a surface operator to each of the $\SU(2)$  gauge groups.
In some sense our finding is a conformal-block version of their mapping, which worked at the level of the correlation functions,
but the details are yet to be filled.

This discussion brings us another question: in the case without the surface operator, one can consider not just the chiral conformal blocks but the correlation function of  Liouville/Toda theory by putting the gauge theory on $S^4$.
There, the physical partition function can be evaluated using Pestun's localization \cite{Pestun:2007rz}, and the one-loop part provided the three-point functions of  Liouville/Toda theory.
Presumably, with the insertion of the full surface operator the one-loop part provides the three-point functions so that the partition function on $S^4$ would realize the $\SL(N)$ Wess-Zumino-Wtten theory.
Again this is a mere speculation and the details need to be filled in.

Finally, the observation that the instanton partition function with the surface operator gives the eigenfunction of the quantized Hamiltonians of the Hitchin system should have a natural interpretation in the framework of \cite{Nekrasov:2010ka}.

\section*{Acknowledgements}

The authors are greatly indebted to H.~Nakajima and A.~Okounkov who kindly and painstakingly explained various mathematical concepts to them.
LFA is supported in part by the DOE grant DE-FG02-90ER40542.
YT is supported in part by the NSF grant PHY-0503584, and by the Marvin L.
Goldberger membership at the Institute for Advanced Study.

\appendix

\section{$SL(2)$ current algebra conformal blocks}\label{appendix}

In this appendix we review some aspects of $SL(2)$ current
algebra and sketch the computation of its conformal blocks by
the sewing method.

\subsection{Preliminaries}

The $\SL(2)_k$ current algebra is spanned by the generators
$J_n^a$, where $a=0,\pm$ and $n~ \epsilon ~\mathbb{Z}$. These
generators satisfy the following commutation relations
\begin{eqnarray}
~[J_n^0,J_m^0]&=&\frac{k}{2}n \delta_{n+m,0},\\
~[J_n^0,J_m^\pm]&=&\pm J_{m+n}^\pm, \\
~[J_n^+,J_m^-]&=&2 J^0_{m+n}+k n \delta_{n+m,0}.
\end{eqnarray}
Note that our convention is that the central charge is \begin{equation}
c=\frac{3k}{k+2}.
\end{equation}

Primary states $\ket{j}$ are defined by the
conditions
\begin{eqnarray}
J_{1+m}^{-} \ket{j} &=&  J_{1+m}^{0} \ket{j} = J_m^{+} \ket{j} = 0,~~~~~(m=0,1,\ldots) \\
J_0^0 \ket{j} &=& j\ket{j}.
\end{eqnarray} Their dimension is given by \begin{equation}
\Delta_j=\frac{j(j+1)}{2(k+2)}.
\end{equation}
Descendants are obtained by applying $J_{1-n}^{-}$, $J_{-n}^+$
and $J_{-n}^0$, with $n=1,...$, to primary states.

Let us denote the corresponding primary fields by $\bV_j(x,z)$.
Here $z$ is the coordinate on the worldsheet and $x$ is the label on which $SL(2)$ acts.
The OPE is given by \begin{equation}
J^a(z) \bV_j(x,w) \sim \frac{1}{z-w} \cD^a \bV_j(x,w)
\end{equation} where the differential operators $\cD^a$ are given by
\begin{eqnarray}
{\cal D}^+&=&-x^2 \partial_x+2 j x,\\
{\cal D}^0&=&-x \partial_x+j,\\
{\cal D}^-&=&-\partial_x.
\end{eqnarray}
In terms of modes, we have
\begin{equation}
\label{action}
~[J_n^a,\bV_j(x,z)]=z^n {\cal D}^a \bV_j(x,z).
\end{equation}

\subsection{The sewing method}

In the following we explain how the sewing procedure can be
used in order to compute conformal blocks for the present case, following \cite{Awata:1992sm}.
The basis for the sewing procedure is the following schematic
expression
\begin{equation}
\langle {\cal O}_1 ... {\cal O}_k {\cal O}_{k+1}...{\cal O}_n \rangle = \sum_{j,\mathbf{m},\mathbf{n}} \langle {\cal O}_1 ... {\cal O}_k {\cal J}_{-\mathbf{m}} \bV_j \rangle (K^{-1})^{\mathbf{mn}}\langle {\cal J}_{-\mathbf{n}} \bV_j{\cal O}_{k+1}...{\cal O}_n \rangle
\end{equation}
which allows to compute conformal blocks.
The indices $\mathbf{m}$, $\mathbf{n}$
denote the descendants of such primary fields, namely
\begin{equation}
{\cal J}_{-\mathbf{m}} \bV_i = J_{-n_1}^{a_1}...J_{-n_m}^{a_m} \bV_i.
\end{equation}
Then $K_{\mathbf{mn}}(j)=\vev{(\cJ_{-\bf m}\bV_j)(\cJ_{-\bf n}\bV_j)}$.

\begin{figure}[h]
\centering
\includegraphics[scale=0.4]{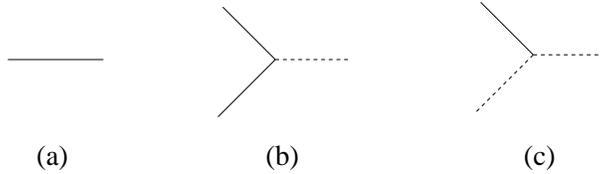}
\caption{Pictorial representation of the propagator $K^{-1}$, shown in (a), the vertex with two external legs $R$, shown in (b) and the vertex with one external leg, shown in (c). External legs are represented with solid lines, while internal legs are represented with dashed lines.\label{sewing}}
\end{figure}

In order to construct the conformal blocks we also need the
triple vertices
\begin{eqnarray}
R_{\bf m}(j_1,j_2,j_3)&=& \frac{\langle {\cal O}_1{\cal O}_2 {\cal J}_{-\bf m} \bV_3  \rangle}{\langle {\cal O}_1{\cal O}_2 \bV_3  \rangle}, \cr
S_{\bf mn}(j_1,j_2,j_3)&=& \frac{\langle {\cal J}_{-\bf m}  \bV_1 {\cal O}_2 {\cal J}_{-\bf n} \bV_3  \rangle}{\langle \bV_1 {\cal O}_2 \bV_3  \rangle}.
\end{eqnarray}
These can be easily obtained by using the action of the
generators on primaries as differential operators
(\ref{action}) and the explicit coordinate dependence of the
three point function of primary fields
\begin{eqnarray}
\vev{\bV^{j_1}(x_1,z_1) \bV^{j_2}(x_2,z_2) \bV^{j_3}(x_3,z_3)} =
\prod_{(abc)} |z_{ab}|^{2(\Delta_{j_c}-\Delta_{j_a}-\Delta_{j_b})}|x_{ab}|^{2(j_a+j_b-j_c)}
\end{eqnarray} where the product on $(abc)$ runs over the cyclic permutations of $(123)$. $x_{ab}=x_a-x_b$ and $z_{ab}=z_a-z_b$.
It is convenient to represent pictorially $K^{-1}$ as a
propagator and $R$ and $S$ as triple vertices. The difference
between the two is the amount of external legs, two for $R$ and
one for $S$, see figure~\ref{sewing}.

\begin{figure}[h]
\centering
\includegraphics[scale=0.4]{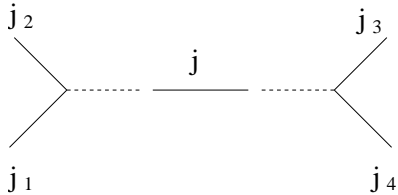}
\caption{Sewing of building blocks into the four point conformal block on the sphere\label{four}}
\end{figure}

Conformal blocks can then be obtained by combining these
building blocks. For instance, the four point conformal block
on the sphere is given by $R(j_4,j_3,j) K^{-1}(j) R(j,j_2,j_1)$
and can be represented pictorially as in figure \ref{four}.
The level of each contribution to a given conformal block is
fixed by the level of its internal propagators, $k_1$, $k_2$,
for the first, etc. The full conformal block is then obtained
by multiplying each contribution by a factor like $z_1^{k_1}
x_1^{k_2}$ per each block. 

\section{Bifundamental contribution}\label{examples}
Here we would like to illustrate how the instanton counting with the full surface operator is performed with a few explicit calculations.
We will stick to the case $N=2$.

As we discussed, the instanton moduli space is denoted by $\cM_{2;k_1,k_2}$.
The fixed points under $\U(1)^2\times \U(1)^N$ are labeled by a pair of Young diagrams $\vec\eth=(\eth_1,\eth_2)$ such that \begin{align}
k_1 &= \sum_{o} \eth_{1,o} + \sum_{e} \eth_{2,e} , &
k_2 &= \sum_{e} \eth_{1,e} + \sum_{o} \eth_{2,o} .
\end{align} Here the summations run over odd and even positive integers, respectively.

For example,
the fixed points in $\cM_{2;1,1}$ are \begin{equation}
(\Young{2},\varnothing),\
(\Young{1},\Young{1}),\
(\varnothing,\Young{2})
\end{equation} and those in
$\cM_{2;2,1}$ are \begin{equation}
(\Young{1} ,\Young{2}),\
(\Young{11} ,\Young{1}),\
(\Young{11}\Young{1} ,\varnothing),\
(\Young{3} ,\varnothing).
\end{equation}
Here and in the following, we use the convention that $\eth_i$ for a Young diagram $\eth$ is the height of the $i$-th column.

Take $p=(\varnothing,\Young{2})\in\cM_{2;1,1}$ and
$q=(\Young{11},\Young{1}) \in \cM_{2;2,1}$.
Then the vector space $\Ext(p,q)$ is five dimensional.
The character of the vector space is then given by  \eqref{character}:
\begin{equation}
e^{-a+b}+e^{a-b+2\epsilon_1}+e^{a+b+2\epsilon_1}
+e^{a+b+\epsilon_1}+e^{a+b+\epsilon_1+\epsilon_2}.
\end{equation} Here $\epsilon_{1,2}$ are two rotation angles of the spacetime, and $(a,-a)$, $(b,-b)$ are the angles associated to $\U(2)$ acting on $\cM_{2;1,1}$ and $\cM_{2;2,1}$, respectively.
Note that the complicated formula \eqref{character} becomes a sum of five exponentials, as it should be.
Then the bifundamental contribution at this pair of fixed points is
\begin{equation}
(-a+b-m)(a-b+2\epsilon_1-m)(a+b+2\epsilon_1-m)
(a+b+\epsilon_1-m)(a+b+\epsilon_1+\epsilon_2-m).
\end{equation}

Then the first few terms of the instanton partition function of the $\cN=2^*$ theory are \begin{multline}
Z=1+\frac{(\epsilon_1-m)(-2a+\epsilon_1-m)}{\epsilon_1(-2a+\epsilon_1)}x
+\frac{(\epsilon_1-m)(2a+\epsilon_1+\epsilon_2-m)}{\epsilon_1(2a+\epsilon_1+\epsilon_2)}y \\
+ \frac{(\epsilon_1-m)(-2a+\epsilon_1-m)(2\epsilon_1-m)(-2a+\epsilon_1-m)}{2\epsilon_1^2(-2a+\epsilon_1)(-2a+2\epsilon_1)} x^2
+ \cdots.
\end{multline}

\section{Free field representations}\label{free}
Here, we summarize very briefly the free-field representations of
$W_N$ algebras and affine $SL(N)$ algebras, mainly to fix the notation.
We use the convention that a single real boson $\varphi(z)$ has the OPE
\begin{equation}
\varphi(z)\varphi(0)\sim -\log z.
\end{equation}
With the background charge $Q$, the central charge is $c=1+12Q^2$, and the operator $\exp[a\varphi(z)]$ is of dimension $a(2Q-a)/2$.

To construct $W_N$ algebra, one uses $N-1$ free bosons $\vec\varphi(z)$ with the background charge $Q\vec\rho$.
Here $\rho$ is the Weyl vector  of $SL(N)$.
The central charge is \begin{equation}
c=(N-1)+12Q^2|\vec\rho|^2 = (N-1)+(N-1)N(N+1)Q^2.
\end{equation}
The Fock space constructed from the primary state $\exp\left[\vec\alpha\cdot\vec\varphi(z)\right]$ is
the Verma module $V_{\vec \alpha}$ of $W_N$ algebra
for generic $\vec\alpha$. Its dimension is \begin{equation}
h=\frac12\vec\alpha\cdot(2Q\vec\rho-\vec\alpha).
\end{equation}
The generators of $W_N$ algebras are given by the quantum Miura transformation of the free bosons $\vec\varphi(z)$.

The free field representation of the affine $SL(N)$ algebra is the Wakimoto representation. We again use $N-1$ free bosons $\vec\varphi(z)$, now with background charge $\vec\rho b$.
In addition, we introduce a $\beta\gamma$ system
$\beta^a(z)$, $\gamma^a(z)$ of spin $(1,0)$ for each positive root.
Then the level $k$ is \begin{equation}
k=-N-b^{-2}
\end{equation} and the central charge is
\begin{equation}
c=\frac{k(N^2-1)}{k+N} = \left[(N-1)+(N-1)N(N+1)b^2\right] + N(N-1).
\end{equation}
The operator $\exp[\vec\alpha\cdot\vec\varphi(z)]$
is again the primary field; the weight vector is $\vec\jmath=-b^{-1}\vec\alpha$. The dimension is then \begin{equation}
h=\frac12\vec\alpha\cdot (2b^{-1}\vec\rho-\vec\alpha)=\frac{\vec \jmath\cdot (\vec \jmath + 2\vec\rho)}{2(k+N)}.
\end{equation}

The Drinfeld-Sokolov reduction involves setting $E^a(z)=1$
for the simple roots $E^a$. To perform this procedure quantum mechanically, one first improves the energy-momentum tensor so that $E^a(z)$ for the simple roots should have dimension zero,
and then introduces $bc$ ghosts to implement the BRST operation.

In the free-field realization, the improvement of the energy-momentum tensor corresponds to the change of the background charge from $b^{-1}\vec\rho$ to $Q\vec\rho=(b+b^{-1})\vec\rho$.
The $bc$ ghosts pair up with the $\beta\gamma$ systems under the BRST operation and they disappear together.
Therefore the Verma module of affine $SL(N)$ with spin $\vec\jmath$
is mapped to the Verma module of W-algebra with momentum
\begin{equation}
\vec\alpha=-b\vec\jmath\label{qDS-momenta}.
\end{equation}


\def\url#1{available at \href{#1}{#1}}
\bibliography{bib}
\bibliographystyle{utphys}

\end{document}